\documentclass[a4paper]{article}

\usepackage{ISCSLP2024}
\usepackage{amsmath,amsfonts}
\usepackage{algorithmic}
\usepackage{algorithm}
\usepackage{array}
\usepackage[caption=false,font=normalsize,labelfont=sf,textfont=sf]{subfig}
\usepackage{textcomp}
\usepackage{stfloats}
\usepackage{url}
\usepackage{verbatim}
\usepackage{graphicx}
\usepackage{cite}

\usepackage{amsmath,mathtools}
\usepackage{tipa}
\usepackage{amssymb}
\usepackage{amsthm}
\usepackage{mathrsfs}
\usepackage{bm, upgreek}
\usepackage[colorlinks = true]{hyperref}
\usepackage{physics}
\usepackage{upgreek}

\usepackage{textcomp}

\usepackage{booktabs}
\usepackage{siunitx}
\usepackage{multirow}
\usepackage{textcomp}
\usepackage{romannum}

\usepackage{xcolor}
\definecolor{BrewerSetRed}{RGB}{228,26,28}
\definecolor{BrewerSetBlue}{RGB}{55,126,184}
\definecolor{BrewerSetGreen}{RGB}{77,175,74}
\definecolor{BrewerSetPurple}{RGB}{152,78,163}
\definecolor{BrewerSetOrange}{RGB}{255,127,0}
\definecolor{BrewerSetYellow}{RGB}{236,176,23}
\definecolor{BrewerSetBrown}{RGB}{166,86,40}
\definecolor{BrewerSetPink}{RGB}{247,129,191}
\definecolor{BrewerSetGray}{RGB}{153,153,153}

\usepackage{tikz}
\usetikzlibrary{shapes.geometric,positioning}
\DeclareRobustCommand\SolidLine[1]{
    \tikz[baseline=-0.6ex]\draw[very thick, #1] (0,0) -- (2em,0);
}

\title{The feasibility of sound zone control using an array of parametric array loudspeakers}
\name{Tao Zhuang$^1$, Jia-Xin Zhong$^2$, Jing Lu$^1$}
\address{
  $^1$Nanjing University, Key Laboratory of Modern Acoustics, Nanjing 210008, China\\
  $^2$The Pennsylvania State University, Graduate Program in Acoustics, University Park, PA 16802, USA}
\email{taozhuang@smail.nju.edu.cn, jiaxin.zhong@psu.edu, lujing@nju.edu.cn}

\begin{document}

\maketitle
\begin{abstract}

Parametric array loudspeakers (PALs) are known for producing highly directional audio beams, a feat more challenging to achieve with conventional electro-dynamic loudspeakers (EDLs).
Due to their intrinsic physical mechanisms, PALs hold promising potential for spatial audio applications such as virtual reality (VR). 
However, the feasibility of using an array of PALs for sound zone control (SZC) has remained unexplored, mainly due to the complexity of the nonlinear demodulation process inherent in PALs.
Leveraging recent advancements in PAL modeling, this work proposes an optimization algorithm to achieve the acoustic contrast control (ACC) between two target areas using a PAL array.
The performance and robustness of the proposed ACC-based SZC using PAL arrays are investigated through simulations, and the results are compared with those obtained using EDL arrays.
The results show that the PAL array outperforms the EDL array in SZC performance and robustness at higher frequencies and lower signal-to-noise ratio, while being comparable under other conditions.
This work paves the way for high-contrast acoustic control using PAL arrays.

\end{abstract}
\noindent\textbf{Index Terms}: parametric array loudspeaker, sound zone control, spatial audio, directional loudspeaker

\section{Introduction}

Parametric array loudspeakers (PALs) have been widely used in many audio applications such as spatial sound reproduction \cite{Tan2012}, virtual sound source construction \cite{Harma2008} and stereo reproduction \cite{Aoki2012} due to their capability of generating highly directional sound beams \cite{Gan2012, Shi2015, Zhong2024}.
Sound zone control (SZC) aims at adjusting the output of a loudspeaker array to brighten or darken a target zone \cite{Druyvesteyn1997}. 
This technology can be applied to aircraft and/or car cabins \cite{Widmark2019, So2019} and for virtual reality (VR) \cite{Murphy2011, Zhang2017}. 
Current research primarily focuses on using electro-dynamic loudspeaker (EDL) arrays for SZC, and there has been no related research on using PAL arrays for SZC. 
Moreover, compared to EDL arrays, PAL arrays offer advantages such as smaller size and flexible phase control \cite{Shinagawa2008, Shi2014, Hahn2021, Zhong2022a}. 
Therefore, this work attempts to achieve SZC using a PAL array.

The transfer function of an EDL array can be written as a linear superposition of the transfer functions between different elements. 
However, the challenge of using an array of PALs for SZC lies in the complexity of the nonlinear demodulation process inherent in PALs.
Specifically, when a PAL generates two intense ultrasonic beams at different frequencies, audio sound is demodulated due to the nonlinear interactions of ultrasound in air \cite{Gan2012, Cervenka2013, Zhong2021}. 
This means that different frequencies of ultrasound radiated by different PAL elements in an array will also demodulate audio sound due to nonlinear effects. 
Consequently, the transfer function of a PAL array is more complex than that of an EDL array, making the optimization problem for SZC using a PAL array significantly more challenging.

This work models the transfer function of the PAL array based on the quasilinear solution of the Westervelt equation, governing the second-order nonlinear wave dynamics \cite{Aanonsen1984}.
An iterative optimization method based on acoustic contrast control (ACC) \cite{Choi2002} method using a PAL array is proposed. 
The feasibility of SZC using a PAL array is then discussed based on the proposed method. 
Numerical results for SZC using a PAL array with transfer function perturbations, based on the proposed method, are presented in Section~\ref{sec:results}.
The validity and robustness of the proposed method are also evaluated by comparing it to SZC using an EDL array.

% \section{Page layout and style}
\section{Theory}

% \subsection{Problem formulation}

\subsection{Audio sound generated by a  PAL array}
\label{sec:audio_press_PAL_array}

\begin{figure}[!t]
\centering
\includegraphics[width = 0.45\textwidth]{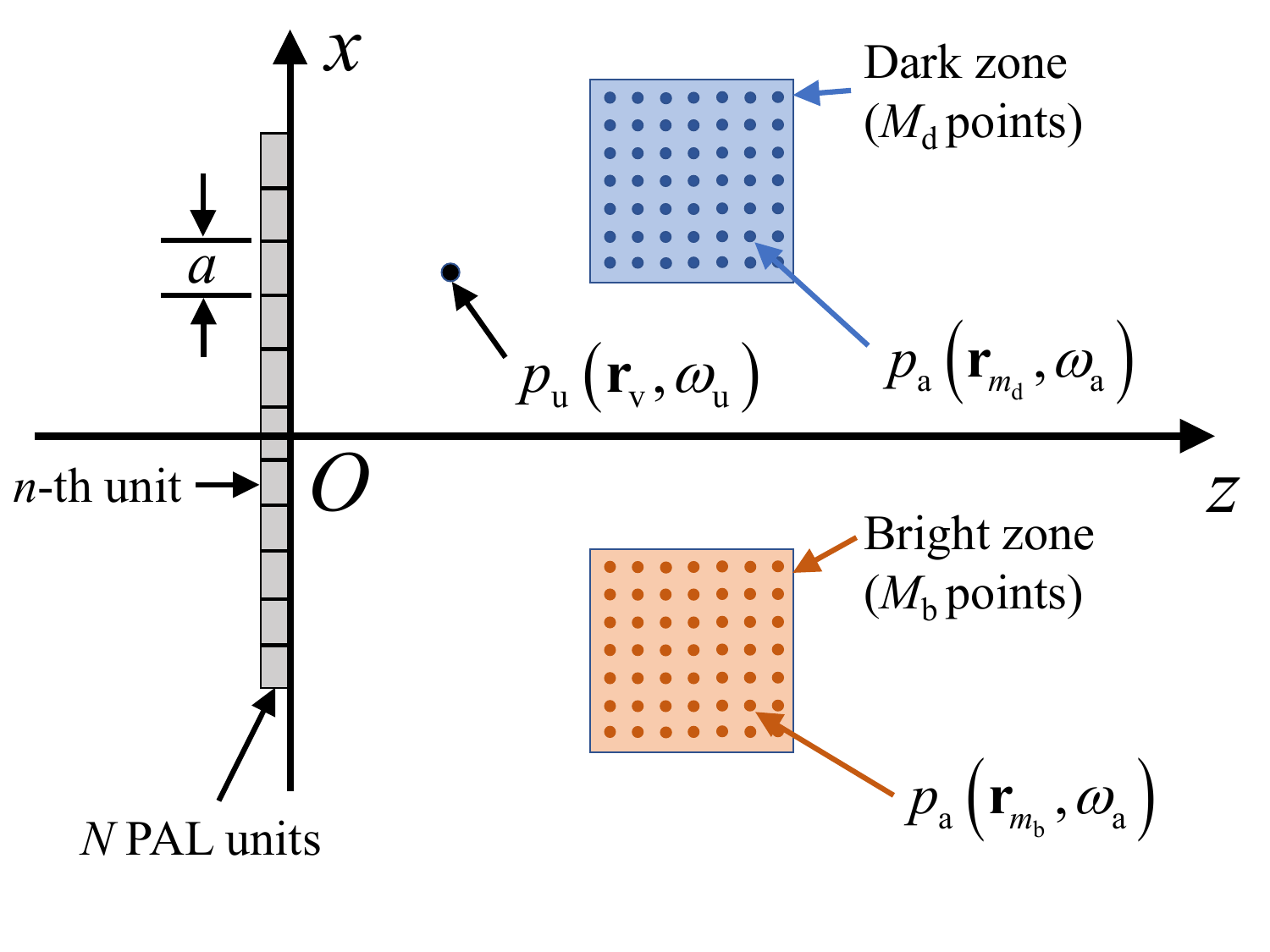}
\caption{\label{fig:sketch_PalArray}{The sketch of a PAL array consisting of $N$ elements. 
Surface vibration velocity of each elements can be modified individually.}}
\end{figure}

As shown in Fig.~\ref{fig:sketch_PalArray}, $N$ rectangular PAL array elements (ultrasonic transducers) are assumed to be located in the plane $Oxy$ with $y$-axis perpendicular to the $Oxz$ plane. 
The rectangular PAL array elements are assumed to be sufficiently long in the $y$-direction, which simplifies the problem to a two-dimensional (2D) condition where the sound field is independent of the $y$.
For simplicity, the width of all PAL elements in the $x$-direction is assumed to be identical of $a$.
Each element generates two harmonic ultrasound waves at frequencies $f_1$ and $f_2$ ($f_1<f_2$). 
In this paper, the subscript u = 1 and 2 represents the index of the ultrasound at frequencies $f_1$ and $f_2$, respectively. 
The angular frequency of the ultrasound is $\omega_\mathrm{u} = 2 \pi f_\mathrm{u}$. 
The vibration velocity on the surface of the $n$-th element at frequency $f_\mathrm{u}$ is denoted as $s_{\mathrm{u}, n}v_0$, where $v_0$ is a constant with the unit of m/s.
Within the framework of the quasilinear solution, the ultrasound pressure at frequency $f_\mathrm{u}$ generated by the PAL array at $\vb{r}_{\mathrm{v}}$ follows the linear acoustic equation \cite{Zhong2024}.
Therefore, it can be expressed as
\begin{equation}
p_\mathrm{u}(\vb{r}_{\mathrm{v}},\omega_\mathrm{u}) 
= \sum_{n = 1}^N {
    s_{\mathrm{u}, n} \hat{p}_{\mathrm{u}, n} \qty(\vb{r}_{\mathrm{v}},\omega_\mathrm{u})
},
\label{eq:PAL_array_ultra_press}
\end{equation}
where the time harmonic component $\exp(-\mathrm{i} \omega_\mathrm{u} t)$ is omitted for simplicity and $\hat{p}_{\mathrm{u}, n} \qty(\vb{r}_{\mathrm{v}},\omega_\mathrm{u})$ represents the ultrasound pressure at frequency $f_\mathrm{u}$ generated by the $n$-th PAL at $\vb{r}_{\mathrm{v}}$ with the a surface vibration velocity of $v_0$, which can be obtained using the Rayleigh integral \cite{Zhong2022}. 

According to the quasilinear solution \cite{Zhong2021}, 
the audio sound radiated by the PAL array can be considered as a superposition of the pressure radiated by infinite virtual audio sources in air with the source density of
\begin{equation}
q (\vb{r}_\mathrm{v},\omega_\mathrm{a}) 
= \frac{\beta \omega_\mathrm{a}} {\mathrm{i} \rho_0^2 c_0^4} 
p_1^* \qty(\vb{r}_\mathrm{v}, \omega_2)
p_2 \qty(\vb{r}_\mathrm{v}, \omega_2)
% \begin{split}
%     q (\vb{r}_\mathrm{v},\omega_\mathrm{a}) 
% & = - \frac{\mathrm{i} \beta \omega_\mathrm{a}} {\rho_0^2 c_0^4} 
% p_1^* \qty(\vb{r}_\mathrm{v}, \omega_2)
% p_2 \qty(\vb{r}_\mathrm{v}, \omega_2) \\ 
% & = - \frac{\mathrm{i} \beta \omega_\mathrm{a}} {\rho_0^2 c_0^4} 
% \sum_{i, j = 1}^N 
%     v_{1,i}^* v_{2,j}
%     \hat{p}_{1, i}^* \qty(\vb{r}_{\mathrm{v}},\omega_1)
%     \hat{p}_{2, j} \qty(\vb{r}_{\mathrm{v}},\omega_2),
%   \label{eq:PAL_array_virtual_source}  
% \end{split}
\label{eq:PAL_array_virtual_source}
\end{equation}
where i is the imaginary unit, $\beta$ is the nonlinearity coefficient, 
the superscript ``*'' denotes the complex conjugate transpose, 
$\rho_0$ is the air density, $c_0$ is the sound velocity in air, $\omega_{\mathrm{a}} = 2 \pi f_\mathrm{a}$, and $f_\mathrm{a} = f_2-f_1$ is the audio frequency. 
Then the audio sound at a field point $\vb{r}_m$ can be obtained by\cite{Zhong2020}
\begin{equation}
 p_\mathrm{a} \qty(\vb{r}_m,\omega_\mathrm{a}) 
  = \frac{\rho_0\omega_\mathrm{a}}{4} 
  \iint_{-\infty}^\infty {
     q \qty(\vb{r}_\mathrm{v},\omega_\mathrm{a}) H_0^{(1)} \qty(k_\mathrm{a} \abs{\vb{r}_m - \vb{r}_\mathrm{v}} )
     \dd[2]{\vb{r}_\mathrm{v}}},
\label{eq:west_audio_press}
\end{equation}
where $k_\mathrm{a} = \omega_\mathrm{a} / c_0$ is the wavenumber of the audio sound and $H_0^{(1)} \qty(\cdot)$ is the Hankel function of the first kind. 

By substituting (\ref{eq:PAL_array_ultra_press}) and (\ref{eq:PAL_array_virtual_source}) into (\ref{eq:west_audio_press}) and exchanging the order of summation and integration, the audio sound generated by the PAL array at a field point $\vb{r}_m$ can be expressed as 
\begin{equation}
% \begin{split}
p_\mathrm{a} \qty(\vb{r}_m,\omega_\mathrm{a}) = 
\vb{s}_1^* \vb{H}_m \vb{s}_2,
\label{eq:PAL_array_audio_press}
\end{equation}
where $\vb{s}_\mathrm{u} = [s_{\mathrm{u}, 1}, s_{\mathrm{u}, 2}, \ldots, s_{\mathrm{u}, N}]^\mathrm{T}$ represents the source vibration velocity vector of the PAL array at frequency $f_\mathrm{u}$, the superscript ``T'' represents the transpose and 
\begin{equation}
    \vb{H}_m = 
    \begin{bmatrix}
        H_{m; 1,1} & H_{m; 1,2} & \cdots & H_{m; 1,N} \\
        H_{m; 2,1} & H_{m; 2,2} & \cdots & H_{m; 2,N} \\
        \vdots & \vdots & \ddots & \vdots \\
        H_{m; N,1} & H_{m; N,2} & \cdots & H_{m; N,N}
    \end{bmatrix}
\end{equation}
represents the transfer function matrix of the audio sound pressure generated by the PAL array, expressed as
\begin{equation}
\begin{split}
    H_{m; i,j}
    & = 
    \frac{\beta\omega_\mathrm{a}^2}{4 \mathrm{i} \rho_0 c_0^4} 
    \iint_{-\infty}^\infty
    \hat{p}_{1,i}^*(\vb{r}_{\mathrm{v}},\omega_1)
    \hat{p}_{2,j}(\vb{r}_{\mathrm{v}},\omega_2) \\
    & \times H_0^{(1)} \qty(k_\mathrm{a} \abs{\vb{r}_m - \vb{r}_\mathrm{v}} )
    \dd[2] \vb{r}_\mathrm{v}.
\end{split}
\end{equation}
Note that when the index $i\neq j$, the physical meaning of $H_{m; i,j} s_{1,i}^* s_{2,j}$ in (\ref{eq:PAL_array_audio_press}) is the audio sound generated by the interaction between the ultrasounds at frequency $f_1$ and $ f_2$, generated by the $i$-th and $j$-th element, respectively, thereby it is denoted as the \emph{coupled audio sound pressure} in this work.
For a special case when $i = j = n$, $H_{m; n,n} s_{1,n}^* s_{2,n}$ in (\ref{eq:PAL_array_audio_press}) represents the audio sound solely generated by the $n$-th element.

Unlike EDL array processing \cite{Han2019},
(\ref{eq:PAL_array_audio_press}) demonstrates that the audio sound generated by an array of PAL elements is not simply the linear superposition of that generated by each element, but it includes the coupled audio sound component.
% Then the audio sound pressure vector of an area contains $M$ field point can be expressed as $\vb{p}_\mathrm{a} = \qty[p_\mathrm{a} \qty(\vb{r}_1,\omega_\mathrm{a}), p_\mathrm{a} \qty(\vb{r}_2,\omega_\mathrm{a}), \ldots, p_\mathrm{a} \qty(\vb{r}_M,\omega_\mathrm{a})]^\mathrm{T}$. 
% Although the audio sound generated by each element can be efficiently obtained using various kinds of methods, e.g., the SWE method \cite{Zhong2020}, 
% the coupled audio sound is still challenging at present and will be solved in the subsequent sections.

\subsection{Acoustic contrast control using a PAL array}

As shown in Fig.~\ref{fig:sketch_PalArray}, the acoustic bright and dark areas contain $M_\mathrm{b}$ and $M_\mathrm{d}$ control points, respectively. 
According to (\ref{eq:PAL_array_audio_press}), the audio sound pressure at $\vb{r}_{m_\mathrm{b}}$ in bright area and $\vb{r}_{m_\mathrm{b}}$ in dark area can be expressed as $p_\mathrm{a,b} ( \vb{r}_{m_\mathrm{b}}, \omega_\mathrm{a} ) = \vb{s}_1^* \vb{H}_{m_\mathrm{b}} \vb{s}_2$ and $p_\mathrm{a,d} ( \vb{r}_{m_\mathrm{d}}, \omega_\mathrm{a} ) = \vb{s}_1^* \vb{H}_{m_\mathrm{d}} \vb{s}_2$, respectively.
Then the audio sound pressure vector of the bright area and the dark area can be expressed as $\vb{p}_\mathrm{a,b} = \qty[p_\mathrm{a,b} \qty(\vb{r}_1,\omega_\mathrm{a}),  p_\mathrm{a,b} \qty(\vb{r}_2, \omega_\mathrm{a}), \ldots, p_\mathrm{a,b} \qty(\vb{r}_{M_\mathrm{b}},\omega_\mathrm{a})]^\mathrm{T}$
and $\vb{p}_\mathrm{a,d} = \qty[p_\mathrm{a,d} \qty(\vb{r}_1,\omega_\mathrm{a}), p_\mathrm{a,d} \qty(\vb{r}_2,\omega_\mathrm{a}), \ldots, p_\mathrm{a,d} \qty(\vb{r}_{M_\mathrm{d}},\omega_\mathrm{a})]^\mathrm{T}$, respectively. 
The cost function of the ACC can be expressed as
\begin{equation}
    \mathcal{J} = \underset{\vb{s}_1, \vb{s}_2}{\max} 
    \frac{\abs{\vb{p}_\mathrm{a, b}}^2}{\abs{\vb{p}_\mathrm{a, d}}^2},
    \label{eq:cost_function_PAL_ACC}
\end{equation}
where $\abs{\cdot}$ is denoted the norm of the vector.
By substituting the expression of the $\vb{p}_\mathrm{a, b}$ and $\vb{p}_\mathrm{a, d}$ into (\ref{eq:cost_function_PAL_ACC}),
the cost function can be expressed as
\begin{equation}
    \mathcal{J} = \underset{\vb{s}_1, \vb{s}_2}{\max} \frac{ \sum_{m_\mathrm{b} = 1}^{M_\mathrm{b}} \vb{s}_2^* \vb{H}_{m_\mathrm{b}}^* \vb{s}_1 \vb{s}_1^* \vb{H}_{m_\mathrm{b}} \vb{s}_2 }{\sum_{m_\mathrm{d} = 1}^{M_\mathrm{d}} \vb{s}_2^* \vb{H}_{m_\mathrm{d}}^* \vb{s}_1 \vb{s}_1^* \vb{H}_{m_\mathrm{d}} \vb{s}_2}.
    \label{eq:cost_function_PAL_ACC_expan1}
\end{equation}
It is hard to obtain the closed solution $\vb{s}_1$ and $\vb{s}_2$ of (\ref{eq:cost_function_PAL_ACC_expan1}) directly. 
However, for a determined $\vb{s}_1$, (\ref{eq:cost_function_PAL_ACC_expan1}) can be expressed as
\begin{equation}
    \mathcal{J} = \underset{\vb{s}_2}{\max} \frac{ \vb{s}_2^* \vb{G}_\mathrm{1,b} \vb{s}_2 }{ \vb{s}_2^* \vb{G}_\mathrm{1,d} \mathrm{\vb{s}_2} },
    \label{eq:cost_function_PAL_ACC_s2}
\end{equation}
where $\vb{G}_\mathrm{1,b} = \sum_{m_\mathrm{b} = 1}^{M_\mathrm{b}} \vb{H}_{m_\mathrm{b}}^* \vb{s}_1 \vb{s}_1^* \vb{H}_{m_\mathrm{b}}$ and $\vb{G}_\mathrm{1,d} = \sum_{m_\mathrm{d} = 1}^{M_\mathrm{d}} \vb{H}_{m_\mathrm{d}}^* \vb{s}_1 \vb{s}_1^* \vb{H}_{m_\mathrm{d}}$. 
Then the optimization problem (\ref{eq:cost_function_PAL_ACC_s2}) has the closed solution which is the eigenvector corresponded to the max eigenvalue of the matrix pair $(\vb{G}_\mathrm{1, b}, \vb{G}_\mathrm{1, d})$.
(\ref{eq:cost_function_PAL_ACC}) can also be expressed as 
\begin{equation}
    \mathcal{J} = \underset{\vb{s}_1, \vb{s}_2}{\max} \frac{ \sum_{m_\mathrm{b} = 1}^{M_\mathrm{b}} \vb{s}_1^* \vb{H}_{m_\mathrm{b}} \vb{s}_2 \vb{s}_2^* \vb{H}_{m_\mathrm{b}}^* \vb{s}_1 }{\sum_{m_\mathrm{d} = 1}^{M_\mathrm{d}} \vb{s}_1^* \vb{H}_{m_\mathrm{d}} \vb{s}_2 \vb{s}_2^* \vb{H}_{m_\mathrm{d}}^* \vb{s}_1 },
    \label{eq:cost_function_PAL_ACC_expan2}
\end{equation}
which is equal to (\ref{eq:cost_function_PAL_ACC_expan1}).
Similarly, for a determined $\vb{s}_2$, (\ref{eq:cost_function_PAL_ACC_expan2}) can be expressed as
\begin{equation}
    \mathcal{J} = \underset{\vb{s}_1}{\max} \frac{ \vb{s}_1^* \vb{G}_\mathrm{2,b} \vb{s}_1 }{ \vb{s}_1^* \vb{G}_\mathrm{2,d} \mathrm{\vb{s}_1} },
    \label{eq:cost_function_PAL_ACC_s1}
\end{equation}
where $\vb{G}_\mathrm{2,b} = \sum_{m_\mathrm{b} = 1}^{M_\mathrm{b}} \vb{H}_{m_\mathrm{b}} \vb{s}_2 \vb{s}_2^* \vb{H}_{m_\mathrm{b}}^*$ and $\vb{G}_\mathrm{2,d} = \sum_{m_\mathrm{d} = 1}^{M_\mathrm{d}} \vb{H}_{m_\mathrm{d}} \vb{s}_2 \vb{s}_2^* \vb{H}_{m_\mathrm{d}}^*$.
Then the solution of (\ref{eq:cost_function_PAL_ACC_s1}) is the eigenvector corresponded to the max eigenvalue of the matrix pair $(\vb{G}_\mathrm{2, b}, \vb{G}_\mathrm{2, d})$.
Therefore, let $N_\mathrm{itr}$ to be the iteration number, the ACC for a PAL array can be solved by an iterative optimization algorithm as shown in Algorithm~\ref{alg:ACC_PAL_array}.
\begin{algorithm}[H]
\caption{ACC for a PAL array.}\label{alg:ACC_PAL_array}
\begin{algorithmic}
\STATE 
\STATE \textbf{calculate} $\vb{H}_\mathrm{b}, \vb{H}_\mathrm{d}$
\STATE \textbf{randomly select} $\vb{s}_1 \subset \mathbb{C}^N $
\STATE \textbf{for} $ i = 1,...,N_\mathrm{itr} $
\STATE \hspace{0.5cm} $\vb{G}_\mathrm{1,b} \gets \sum_{m_\mathrm{b} = 1}^{M_\mathrm{b}} \vb{H}_{m_\mathrm{b}}^* \vb{s}_1 \vb{s}_1^* \vb{H}_{m_\mathrm{b}} $
\STATE \hspace{0.5cm} $\vb{G}_\mathrm{1,d} \gets \sum_{m_\mathrm{d} = 1}^{M_\mathrm{d}} \vb{H}_{m_\mathrm{d}}^* \vb{s}_1 \vb{s}_1^* \vb{H}_{m_\mathrm{d}} $
\STATE \hspace{0.5cm} $\lambda_2 \gets $ max eigenvalue of $\qty(\vb{G}_\mathrm{1,b}, \vb{G}_\mathrm{1,d})$ 
\STATE \hspace{0.5cm} $\vb{s}_2 \gets $ eigenvector corresponded to $\lambda_2$
\STATE \hspace{0.5cm} $\vb{G}_\mathrm{2,b} \gets \sum_{m_\mathrm{b} = 1}^{M_\mathrm{b}} \vb{H}_{m_\mathrm{b}} \vb{s}_2 \vb{s}_2^* \vb{H}_{m_\mathrm{b}}^* $
\STATE \hspace{0.5cm} $\vb{G}_\mathrm{2,d} \gets \sum_{m_\mathrm{d} = 1}^{M_\mathrm{d}} \vb{H}_{m_\mathrm{d}} \vb{s}_2 \vb{s}_2^* \vb{H}_{m_\mathrm{d}}^* $
\STATE \hspace{0.5cm} $\lambda_1 \gets $ max eigenvalue of $\qty(\vb{G}_\mathrm{2,b}, \vb{G}_\mathrm{2,d})$ 
\STATE \hspace{0.5cm} $\vb{s}_1 \gets $ eigenvector corresponded to $\lambda_1$
\STATE \textbf{normalize}  $\vb{s}_1, \vb{s}_2$
\STATE \textbf{return}  $\vb{s}_1, \vb{s}_2$
\end{algorithmic}
\end{algorithm}

\section{Results\label{sec:results}}

\subsection{Parameters setting}

In this section, the PAL array consists of 24 elements, closely arranged along the $x$-axis and the center of the PAL array is coincided with $O$. 
The width of each element of the PAL array is $a = 1\,\mathrm{cm}$. 
To fairly compare the performance of the PAL array with that of EDL arrays, the EDL array is set to the same dimensions as the PAL array. 
However, it is important to note that, in practice, EDL arrays cannot achieve such compact sizes as assumed under simulation conditions. 
Hereafter, the bright zone is set as a rectangular area of $-0.6 \text{ m} \leq x \leq -0.3 \text{ m}$, $0.6 \text{ m} \leq z \leq 0.9 \text{ m}$ with $10 \times 10$ control points uniformly distributed for calculating the transfer function $\vb{H}_{m_\mathrm{b}}$; the dark zone is set as a rectangular area of $0.3 \text{ m} \leq x \leq 0.6 \text{ m}$, $0.6 \text{ m} \leq z \leq 0.9 \text{ m}$ with the same $10 \times 10$ control points uniformly distributed for calculating the transfer function $\vb{H}_{m_\mathrm{d}}$. 
The acoustic contrast is defined as
\begin{equation}
    \gamma_\mathrm{AC} = 10 \log_{10} \qty( E_\mathrm{bright} / E_\mathrm{dark} ),
\end{equation}
where $E_\mathrm{bright}$ and $E_\mathrm{dark}$ is the audio sound energy in the bright and dark zone, respectively.
Simulations are conducted under free-field conditions. 
To simulate the impact of sound field perturbations on the performance of different algorithms, a perturbation is added to the transfer function before using the optimization algorithm, shown as
\begin{equation}
    \widetilde{H}_m = 
    \qty( \abs{H_m} + U_\mathrm{amp} )
    \exp \qty[\mathrm{i}\qty( \angle{H_m} + U_\mathrm{phase} )]
\end{equation}
where $U_\mathrm{amp}$ is the amplitude perturbation which is Gaussian noise with a signal-to-noise ratio (SNR) and $U_\mathrm{phase} = R \cdot \mathcal{N}(0,1)$ is the phase perturbation, where $\mathcal{N}(0,1)$ is the standard normal distribution and $R$ is a constant denoted as the range of the phase perturbation. 
Other simulation parameters are listed in Table~\ref{tab:para_setting}.

\begin{table}[th]
  \caption{Parameters setting}
  \label{tab:para_setting}
  \centering
  \begin{tabular}{ r@{}l  r }
    \toprule
    & \multicolumn{1}{c}{\textbf{Parameters}} & 
    \multicolumn{1}{c}{\textbf{Values}} \\
    \midrule
    & Center frequency of the ultrasound & 40~kHz\\
    & Temperature of the air & 20~$^\circ$C\\
    & Humidity of the air & 70~\%\\
    & Density of the air & 1.21~kg/$\mathrm{m}^3$\\
    & Velocity of the sound & 343~m/s\\
    & Nonlinear coefficient & 1.2\\
    \bottomrule
  \end{tabular}
\end{table}

\subsection{Convergence analysis}

\begin{figure}[!t]
\centering
\includegraphics[width = 0.35\textwidth]{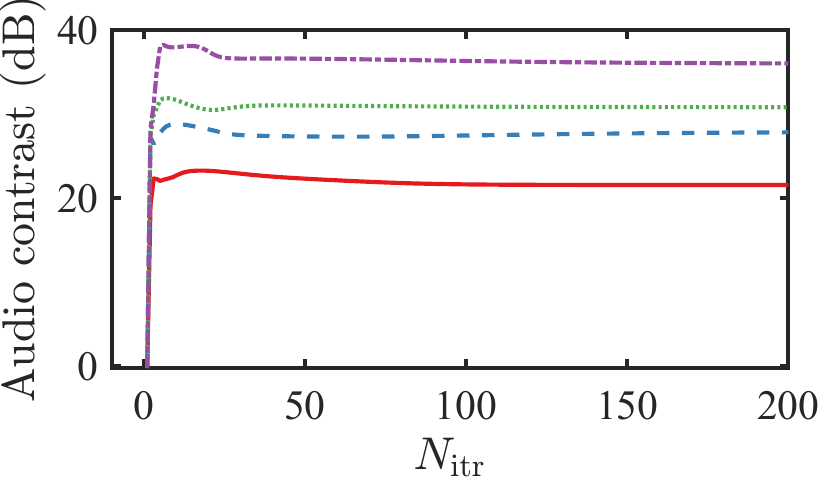}
\caption{\label{fig:Conver_PalArray}{The acoustic contrast by ACC using the PAL array as the number of iteration number $N_\mathrm{itr}$ increases with audio frequency: \SolidLine{color=BrewerSetRed}, 1~kHz; \SolidLine{color=BrewerSetBlue, dashed}, 2~kHz; \SolidLine{color=BrewerSetGreen, densely dotted}, 4~kHz and \SolidLine{color=BrewerSetPurple, dash dot}, 8~kHz. 
The SNR of the amplitude perturbation is set as 30~dB and the range of the phase perturbation is set as 15$^\circ$.
% The sound field perturbations are set as 0.
}}
\end{figure}

To analyze the convergence of the proposed optimization method for the ACC using PAL array, the acoustic contrast of the audio sound field after ACC using the PAL array as discussed in Sec.~\ref{sec:audio_press_PAL_array} related to the iteration number is shown in Fig.~\ref{fig:Conver_PalArray}. 
It can be observed that the acoustic contrast increases with the number of iterations and converges when the number of iterations is sufficiently large ($N_\mathrm{itr} > 100$ in this case). 
This result verifies the convergence of the proposed algorithm and preliminarily confirms that the proposed method can improve the acoustic contrast between the bright and dark zones. 
In subsequent simulations, the number of iterations $N_\mathrm{itr}$ is set to 200 to ensure the convergence of the proposed method.

\subsection{Audio sound field by the ACC using the PAL array}

\begin{figure}[!t]
\centering
\includegraphics[width = 0.45\textwidth]{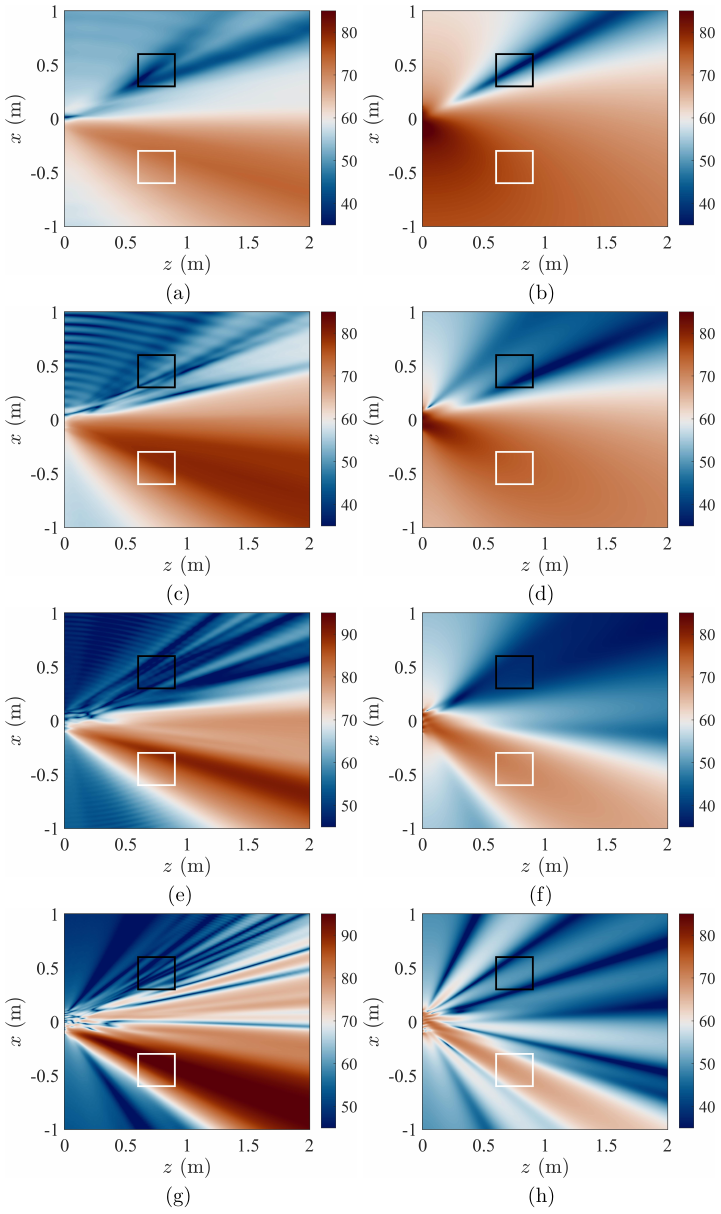}
\caption{\label{fig:Audio_sound_field_ACC}{The audio sound field in sound pressure level (dB, re 20\,$\mu$Pa) by ACC using the PAL array (left column) and the EDL array (right column). The audio sound frequency from top row to bottom row is 1~kHz, 2~kHz, 4~kHz and 8~kHz. The area surrounded by the white and black rectangles are the bright and dark zones, respectively. 
The SNR of the amplitude perturbation is set as 30~dB and the range of the phase perturbation is set as 15$^\circ$.}
}
\end{figure}

Figure~\ref{fig:Audio_sound_field_ACC} shows the audio sound fields radiated in the $Oxz$ plane after ACC optimization using both PAL and EDL arrays. 
It can be observed that the proposed ACC method based on PAL arrays effectively achieves sound field zoning with acoustic contrast 25.3~dB, 31.0~dB, 34.2~dB and 36.8~dB at 1~kHz, 2~kHz, 4~kHz and 8~kHz, while the acoustic contrast of the EDL array is 22.4~dB, 31.1~dB, 24.0~dB and 18.1~dB.
% Additionally, it can be seen that the performance of sound field zoning is superior when the audio frequency is above 4 kHz compared to that of the EDL array. 
This is because, compared to the direct radiation of audio sound by an EDL array, the resolution of the audio sound field radiated by a PAL array is determined by the carrier ultrasound \cite{Zhong2022a}. 
Since ultrasound has a smaller wavelength than audio sound, the PAL array can achieve better spatial resolution for the audio sound.
% This is because the PAL array has stronger directivity than the EDL array, which results in better concentration of sound energy. 
This can be verified by observing Figure~\ref{fig:Audio_sound_field_ACC}, which shows that in the region close to the $x$-axis, the audio sound pressure radiated by the PAL array is significantly lower than the sound field radiated by the EDL array in the same region.

\subsection{Robustness analysis}
To analyze and compare the robustness of SZC using the PAL array and the EDL array, for each sound field perturbation decided by  the SNR of the amplitude perturbation and the range of the phase perturbation, 100~perturbations are applied to the transfer function, and SZC is performed using both the PAL array and the EDL array for each perturbed transfer function. 
The resulting acoustic contrast data is then statistically analyzed. 

\begin{figure}[!t]
\centering
\includegraphics[width = 0.45\textwidth]{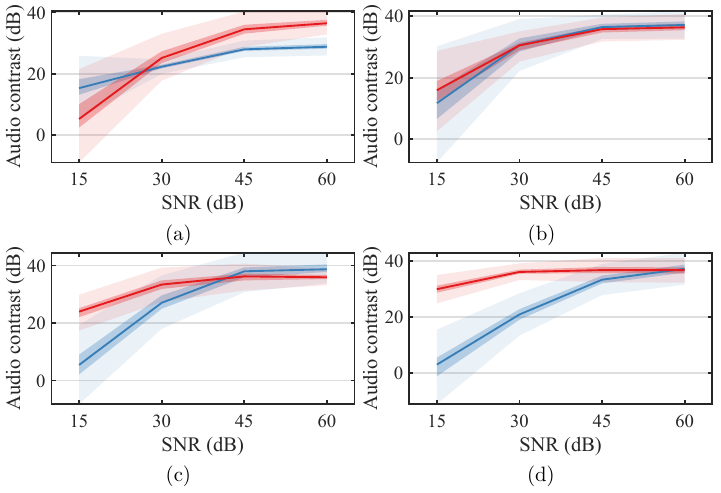}
\caption{\label{fig:Robust_SNR}{The acoustic contrast of the PAL array (red line) and the EDL array (blue line) at different SNR of the amplitude perturbation. The range of phase perturbation is $15^\circ$. The audio frequency is 1~kHz (a), 2~kHz (b), 4~kHz (c) and 8~kHz (d). }}
\end{figure}

Figure~\ref{fig:Robust_SNR} illustrates the robustness of the PAL array and the EDL array to amplitude perturbations in the transfer function.
% with the phase perturbation range fixed at 15$^\circ$ and the amplitude perturbation SNR set to 15~dB, 30~dB, 45~dB and 60~dB, respectively.
It can be observed that when the audio frequency is above 2~kHz and the SNR is larger than 45~dB, the acoustic contrast of the PAL array is close to that of the EDL array. 
However, as the SNR decreases, the acoustic contrast of the PAL array is significantly higher than that of the EDL array, indicating that the PAL array has better robustness at high frequencies compared to the EDL array. 
This is because the EDL array has poor directivity at high frequencies, which causes perturbations at the control points to lead to sound energy leakage, resulting in a decline in SZC performance. 
In contrast, the PAL array has strong directivity and high-frequency response, so even with perturbations at the control points, the sound energy remains concentrated in the bright zone, leading to higher acoustic contrast compared to the EDL array.

\begin{figure}[!t]
\centering
\includegraphics[width = 0.45\textwidth]{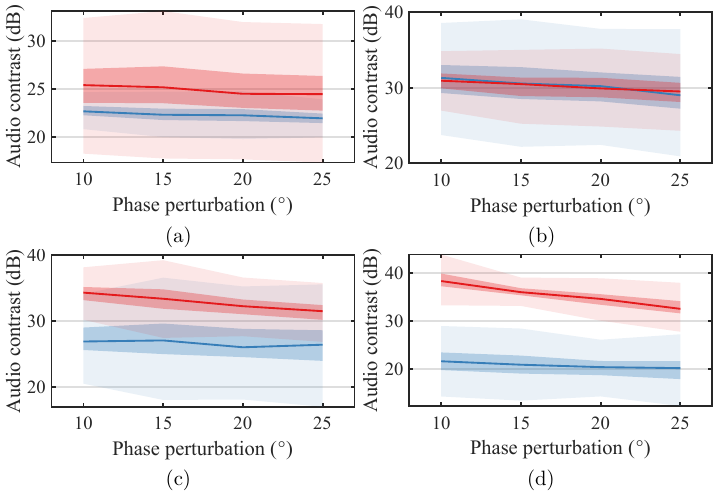}
\caption{\label{fig:Robust_phase}{The acoustic contrast of the PAL array (red line) and the EDL array (blue line) at different range of the phase perturbation. The SNR of the amplitude perturbation is 30~dB. The audio frequency is 1~kHz (a), 2~kHz (b), 4~kHz (c) and 8~kHz (d).}}
\end{figure}

Figure~\ref{fig:Robust_phase} illustrates the robustness of the PAL array and the EDL array to phase perturbations in the transfer function. 
It can be observed that as the phase perturbation range increases, the acoustic contrast for both the PAL array and the EDL array decreases, and the extent of this decrease is similar for both arrays. 
The simulation results indicate that for small-angle (less than $25^\circ$) phase perturbations in the transfer function, the robustness of the PAL array and the EDL array is comparable. 
At this point, the difference in robustness for SZC between the PAL array and the EDL array is mainly affected by the amplitude perturbations of the transfer function.

\section{Conclusions}

In this work, a method based on the quasilinear solution of the Westervelt equation and the ACC method was proposed to achieve SZC between two target zones using a PAL array.
Simulation results confirm the convergence and feasibility of the proposed optimization algorithm. 
It was observed that the acoustic contrast between the bright and dark zones was significantly enhanced using the PAL array. 
Furthermore, the SZC performance of the PAL array is comparable to that of the EDL array at low frequencies.
Moreover, when operating at high frequencies and under strong sound field disturbances, the PAL array demonstrates superior SZC performance and robustness compared to the EDL array. 
This is due to the PAL's better high-frequency response and strong directivity, which allows for more effective concentration of sound energy in the bright zone, thereby reducing sound energy leakage.
This work has paved the way for simulating and designing of SZC systems using PAL arrays, enabling more robust acoustic control within target areas compared to the EDL array.
Experiments will be conducted in the future to validate the results of this work.
The approaches and results presented in this work may inspire innovative applications in binaural sound reproduction for VR conferences through the use of PAL arrays.

\section{Acknowledgements}
T. Z. and J. L. gratefully acknowledge the financial support by National Natural Science Foundation of China (12274221).

\bibliographystyle{IEEEtran}

\bibliography{mybib}

\end{document}